\documentclass{PoS}

\title{Phase Transition Signature Results from PHENIX}

\ShortTitle{Phase Transition Signature Results from PHENIX}

\author{\speaker{J.T. Mitchell}\thanks{and the PHENIX Collaboration}\\
        Brookhaven National Laboratory\\
        E-mail: \email{mitchell@bnl.gov}}

\abstract{The PHENIX experiment has conducted searches for the QCD critical point with measurements of multiplicity fluctuations, transverse momentum fluctuations, event-by-event kaon-to-pion ratios, elliptic flow, and correlations. Measurements have been made in several collision systems as a function of centrality and transverse momentum.  The results do not show significant evidence of critical behavior in the collision systems and energies studied, although several interesting features are discussed.}

\FullConference{5th International Workshop on Critical Point and Onset of Deconfinement - CPOD 2009,\\
		 June 08 - 12 2009\\
		 Brookhaven National Laboratory, Long Island, New York, USA}

\begin{document}

\section{Introduction}

Recent work with lattice gauge theory simulations indicate that the phase diagram of Quantum Chromodynamics (QCD) may contain a first-order transition line between the hadron gas phase and the strongly-coupled Quark-Gluon Plasma (sQGP) phase that terminates at a critical point \cite{QCDstephanov}. This property is observed in many common liquids, including water. Near the QCD critical point, several thermodynamic properties of the system will diverge with a power law behavior in the variable $\epsilon = (T-T_{C})/T_{C}$, where $T_{C}$ is the critical temperature.  Here, several measurements made by the PHENIX experiment at Brookhaven National Laboratory's Relativistic Heavy Ion Collider that may be sensitive to this critical behavior are discussed.

\section{Multiplicity Fluctuations}

In the Grand Canonical Ensemble, the variance and the mean of the particle number, N, can be directly related to the compressibility, $k_{T}$: $\omega_{N} = \frac{var(N)}{N} = k_B T \frac{N}{V} k_T$, where $k_{B}$ is Boltzmann's constant, T is the temperature, and V is the volume \cite{Stanley}. Near the critical point, the compressibility diverges with a power law behavior with exponent $\gamma$: $k_{T} \propto \epsilon^{-\gamma}$.  The measurement of event-by-event fluctuations in the multiplicity of charged hadrons may be sensitive to critical behavior in the system. PHENIX has surveyed the behavior of inclusive charged particle multiplicity fluctuations as a function of centrality and transverse momentum in $\sqrt{s_{NN}}$=62.4 GeV and 200 GeV Au+Au collisions, and in $\sqrt{s_{NN}}$=22.5, 62.4, and 200 GeV Cu+Cu collisions.

Since multiplicity fluctuations are well described by Negative Binomial Distributions (NBD) in both elementary \cite{ua5} and heavy ion collisions \cite{e802MF}, the data for a given centrality and $p_{T}$ bin are fit to an NBD from which the mean and variance are determined.  Due to the finite width of each centrality bin, there is a non-dynamic component of the observed fluctuations that is present due to fluctuations in the impact parameter within a centrality bin. The magnitude of this component is estimated using the HIJING event generator \cite{HIJING}, which well reproduces the mean multiplicity of RHIC collisions \cite{phxMult}. The estimate is performed by comparing fluctuations from simulated events with a fixed impact parameter to events with a range of impact parameters covering the width of each centrality bin, as determined from Glauber model simulations. The data are corrected to remove the impact parameter fluctuation component. 

\begin{figure}[htb]
                 \resizebox{0.5\textwidth}{!}{%
                 \includegraphics{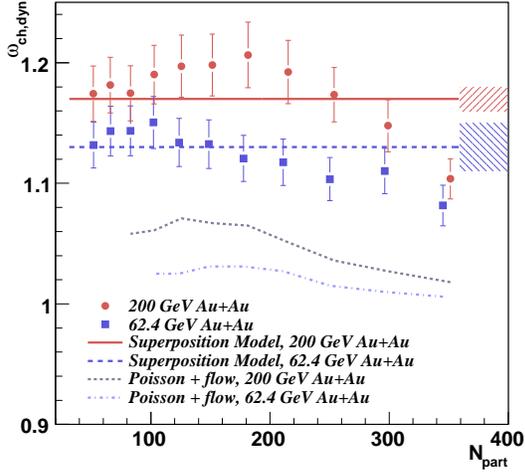}
}
\caption[]{Multiplicity fluctuations as a function of $N_{part}$ for Au+Au collisions for $0.2<p_T<2.0$ GeV/c. Contributions from impact parameter fluctuations have been removed. Shaded regions represent a 1$\sigma$ range of the superposition model prediction derived from p+p data.}
\label{fig:auauSvar}
\end{figure} 

Baseline comparisons are made to the participant superposition model, in which the total multiplicity fluctuations can be expressed in terms of the scaled variance \cite{heiselReview}, $\omega_{N} = \omega_{\nu} + \mu_{WN}~\omega_{N_{part}}$, where $\omega_{\nu}$ are the fluctuations from each individual source, $\omega_{N_{part}}$ are the fluctuations of the number of sources, and $\mu_{WN}$ is the mean multiplicity per wounded nucleon. The second term includes non-dynamic contributions from impact parameter fluctuations along with additional fluctuations in the number of participants for a fixed impact parameter. Ideally, the second term is nearly nullified after applying the previously described corrections, so the resulting fluctuations are independent of centrality as well as collision species. Baseline comparisons at 200 GeV are facilitated by PHENIX measurements of charged particle multiplicity fluctuations in minimum bias 200 GeV p+p collisions with mean $\mu$ = 0.32 $\pm$ 0.003, scaled variance $\omega$ = 1.17 $\pm$ 0.01, and NBD fit parameter $k_{NBD}$ = 1.88 $\pm$ 0.01. 

The scaled variance as a function of the number of participating nucleons, $N_{part}$, over the $p_T$ range $0.2<p_T<2.0$ GeV/c is shown in Figure \ref{fig:auauSvar} for Au+Au collisions. For all centralities, the scaled variance values consistently lie above the Poisson distribution value of 1.0. In all collision systems, the minimum scaled variance occurs in the most central collisions and then begins to increase as the centrality decreases. A similar centrality-dependent trend of the scaled variance has also been observed at the SPS in low energy Pb+Pb collisions at $\sqrt{s_{NN}}$=17.3 GeV, measured by experiment NA49 \cite{na49MF}, where the hard scattering contribution is expected to be small.  All of the data points are consistent with or below the participant superposition model estimate. This suggests that the data do not show any indications of the presence of a critical point, where the fluctuations are expected to be much larger than the participant superposition model expectation.

The clan model \cite{clanOriginal} has been developed to interpret the fact that Negative Binomial Distributions describe charged hadron multiplicity distributions in elementary and heavy ion collisions. In this model, hadron production is modeled as independent emission of a number of hadronic clusters, $N_c$, each with a mean number of hadrons, $n_c$. The independent emission is described by a Poisson distribution with an average cluster, or clan, multiplicity of $\bar{N_c}$. After the clusters are emitted, they fragment into the final state hadrons.  The measured value of the mean multiplicity, $\mu_{\rm ch}$, is related to the cluster multiplicities by $\mu_{\rm ch} = \bar{N_c}\bar{n_c}$. In this model, the cluster multiplicity parameters can be simply related to the NBD parameters of the measured multiplicity distribution as follows:

\begin{equation} \label{eq:clan1}
   \bar{N_c} = k_{\rm NBD}~log(1 + \mu_{\rm ch}/k_{\rm NBD})
\end{equation}

and

\begin{equation} \label{eq:clan2}
   \bar{n_c} = (\mu_{\rm ch}/k_{\rm NBD})/log(1+\mu_{\rm ch}/k_{\rm NBD}).
\end{equation}

\begin{figure}
                 \resizebox{0.5\textwidth}{!}{%
		   \includegraphics{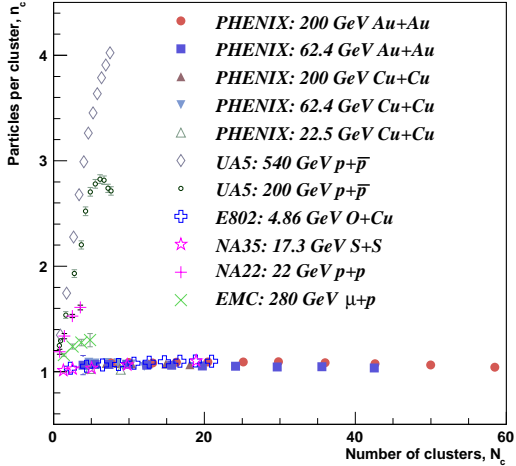}
}
\caption{\label{fig:clanClusAll}
The correlation of the clan model parameters $\bar{n_c}$ and $\bar{N_c}$ for all of the collision species measured as a function of centrality. Also shown are results from pseudorapidity-dependent studies from elementary collisions (UA5~\protect\cite{ua5}, EMC \protect\cite{emcClan}, and NA22~\protect\cite{na22Clan}) and heavy ion collisions (E802~\protect\cite{e802MF} and NA35~\protect\cite{na35Clan}). }
\end{figure}

The results from the NBD fits to the data are plotted in Fig.~\ref{fig:clanClusAll} for all collision species. Also shown are data from elementary and heavy ion collisions at various collision energies. The individual data points from all but the PHENIX data are taken from multiplicity distributions measured over varying ranges of pseudorapidity, while the PHENIX data are taken as a function of centrality. The characteristics of all of the heavy ion data sets are the same. The value of $\bar{n_c}$ varies little within the range 1.0-1.1.  The heavy ion data universally exhibit only weak clustering characteristics as interpreted by the clan model. There is also no significant variation seen with collision energy.  However, $\bar{n_c}$ is consistently significantly higher in elementary collisions. In elementary collisions, it is less probable to produce events with a high multiplicity, which can reveal rare sources of clusters such as jet production or multiple parton interactions.

\section{$\langle p_{T} \rangle$ Fluctuations}

PHENIX has also completed a survey that expands upon previous measurements of event-by-event transverse momentum fluctuations \cite{phxPtFluc}. Here, the magnitude of the $p_{T}$ fluctuations will be quoted using the variable $\Sigma_{p_T}$, as described in \cite{ceresPTF}. $\Sigma_{p_T}$ is the mean of the covariance of all particle pairs in an event, normalized by the inclusive mean $p_T$. $\Sigma_{p_T}$ is related to the inverse of the heat capacity of the system \cite{Korus}, which diverges with a power law behavior near the critical point: $C_{V} \propto \epsilon^{-\alpha}$.

\begin{figure}[htb]
                 \resizebox{0.5\textwidth}{!}{%
                 \includegraphics{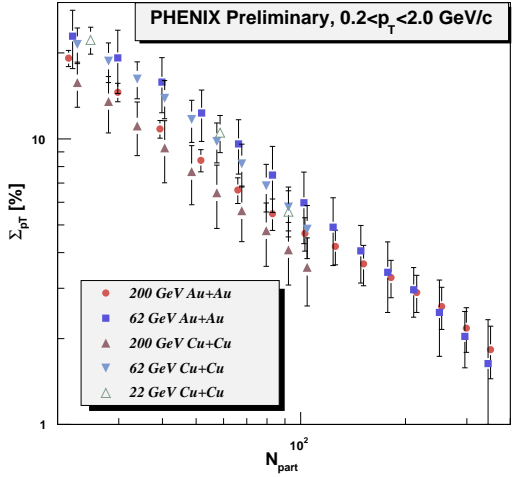}
}
\caption[]{Event-by-event $p_T$ fluctuations for inclusive charged hadrons within the PHENIX acceptance in the transverse momentum range $0.2<p_T<2.0$ GeV/c in terms of $\Sigma_{p_T}$ as a function of $N_{part}$.}
\label{fig:ptfluc}
\end{figure}

Figure \ref{fig:ptfluc} shows $\Sigma_{p_T}$ as a function of $N_{part}$ for all 5 collision systems measured over the $p_T$ range $0.2<p_T<2.0$ GeV/c. The data is shown within the effective PHENIX azimuthal acceptance of 4.24 radians. The magnitude of $\Sigma_{p_T}$ exhibits little variation for the different collision energies and does not scale with the jet cross section at different energies, hence hard processes are not the primary contributor to the observed fluctuations. Simulations show that elliptic flow contributes little \cite{phxPtFluc}. With the exception of the most peripheral collisions, all systems exhibit a universal power law scaling as a function of $N_{part}$. The data points for all systems are best described by the curve: $\Sigma_{p_T} \propto N_{part}^{-1.02 \pm 0.10}$. The observed scaling is independent of the $p_T$ range over which the measurement is made.

\section{K/$\pi$ Fluctuations}

PHENIX has studied identified particle fluctuations by measuring the event-by-event fluctuations of kaons to pions and protons to pions.  One advantage of particle ratio measurements is that contributions from volume fluctuations cancel. Measurements are quoted in the variable $\nu_{dyn}$:
\begin{equation}\label{eq:ktopi}
  \nu_{dyn}(K,\pi) = \frac{\langle \pi(\pi-1)\rangle}{\langle \pi \rangle ^2} + \frac{\langle K(K-1) \rangle}{\langle K \rangle ^2} - 2\frac{\langle K\pi \rangle}{\langle K \rangle \langle \pi \rangle}.
\end{equation}
If only random fluctuations are present, $\nu_{dyn}$ is zero.  Also, $\nu_{dyn}$ is independent of acceptance.

\begin{figure}[htb]
                 \resizebox{0.5\textwidth}{!}{%
                 \includegraphics{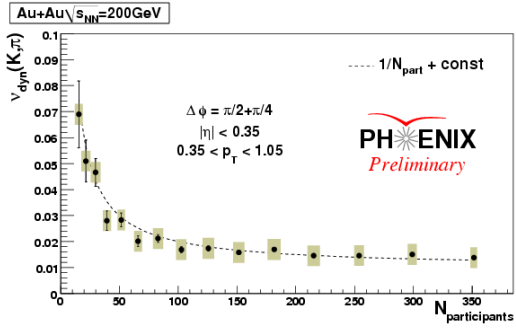}
}
\caption[]{Event-by-event fluctuations of the kaon-to-pion ratio for inclusive charged hadrons within the PHENIX acceptance in the transverse momentum range $0.35<p_T<1.05$ GeV/c. The dashed line is a fit to the function c+$N_{part}^{-1}$, where c is a constant.}
\label{fig:kpifluc}
\end{figure}

\begin{figure}[htb]
                 \resizebox{0.5\textwidth}{!}{%
                 \includegraphics{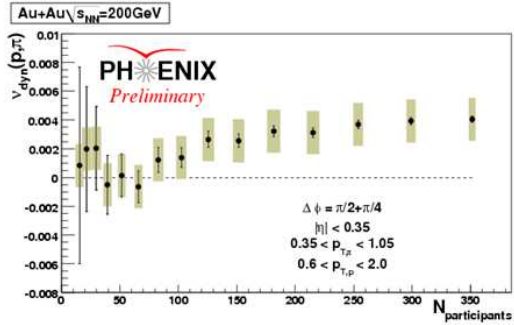}
}
\caption[]{Event-by-event fluctuations of the kaon-to-proton ratio for inclusive charged hadrons within the PHENIX acceptance in the transverse momentum range $0.35<p_T<1.05$ GeV/c.}
\label{fig:kpfluc}
\end{figure}

The measurements for $\nu_{dyn}(K,\pi)$ for $0.34<p_T<1.05$ GeV/c are shown in Figure \ref{fig:kpifluc}.  The measurements for $\nu_{dyn}(K,p)$ are shown in Figure \ref{fig:kpfluc}. As with the $p_T$ fluctuations, the fluctuations in $\langle K \rangle /\langle \pi \rangle$ demonstrate a 1/$N_{part}$ dependence. This is not seen in fluctuations of $\langle p \rangle / \langle \pi \rangle$, which instead rise as centrality increases.
 
\section{Scaling of Elliptic Flow}

\begin{figure}[htb]
                 \resizebox{0.5\textwidth}{!}{%
                 \includegraphics{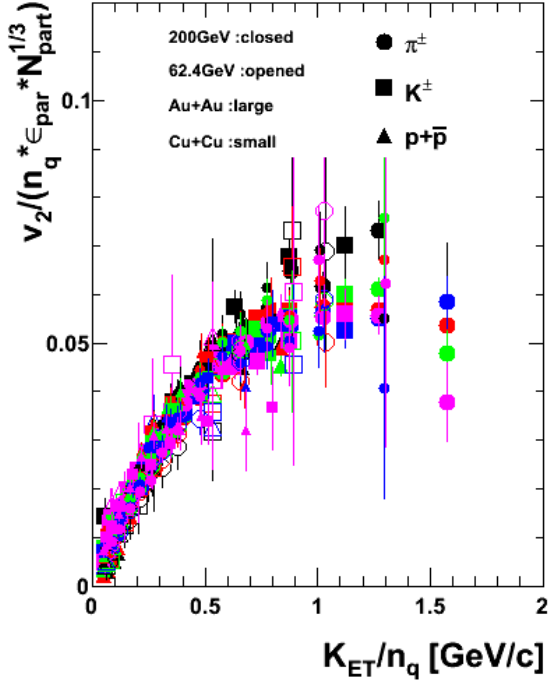}
}
\caption[]{PHENIX Preliminary elliptic flow $v_{2}$ normalized by the number of quarks, collision eccentricity, and $N_{part}^{1/3}$ plotted as a function of the transverse kinetic energy normalized by the number of quarks. Shown are $v_2$ measurements of pions, kaons, and protons over a centrality range of 0-50\% for 200 GeV Au+Au, 200 GeV Cu+Cu, and 62.4 GeV Au+Au.}
\label{fig:v2Scaling}
\end{figure}

One of the most striking RHIC results has been the observation of scaling behavior in elliptic flow measurements below a transverse momentum of about 1 GeV that indicate that quark degrees of freedom are driving the dynamics of the collision \cite{phxFlow}.  PHENIX measurements of the scaling behavior of elliptic flow are compiled for various particle species and various collision systems in Figure \ref{fig:v2Scaling}.  Further measurements of this scaling behavior and the observation of its breaking as a function of collision energy will be an important ingredient in the search for a critical point.

\section{Searching for a Critical Point with HBT Correlations}

\begin{figure}[htb]
                 \resizebox{0.5\textwidth}{!}{%
                 \includegraphics{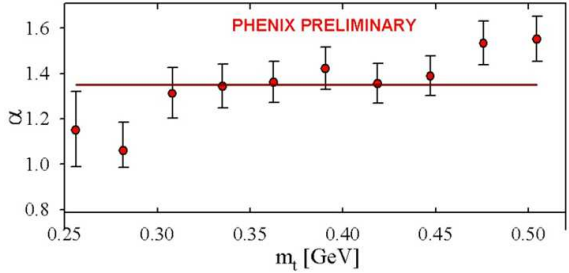}
}
\caption[]{The L\'{e}vy index of stability $\alpha$ extracted from L\'{e}vy function fits to $Q_{inv}$ correlations in 0-5\% central 200 GeV Au+Au collisions as a function of transverse mass.}
\label{fig:hbtAlpha}
\end{figure}

Near the critical point, correlation functions are also expected to be described by a power law function with critical exponent $\eta$. This exponent can be measured with Hanbury-Brown Twiss correlations in the $Q_{inv}$ variable \cite{Csorgo05}. Here, the $Q_{inv}$ correlations are fit with a L\'{e}vy function,
\begin{equation}
  C(Q_{inv}) = \lambda exp(-|Rq/hc|^{-\alpha}),
\end{equation}
where R is the HBT radius and $\alpha$ is the L\'{e}vy index of stability. The value of $\alpha$ is 1 for a Lorentzian source and 2 for a Gaussian source. Since $\alpha$ equates to the exponent $\eta$, it is expected that its value will approach the value expected for the universality class of QCD.  If QCD belongs to the 3d Ising model class, the value of $\eta$ would approach 0.5.  Figure \ref{fig:hbtAlpha} shows the results of the L\'{e}vy function fit to PHENIX HBT correlations in 0-5\% central 200 GeV Au+Au collisions as a function of transverse mass. The fit results are inconsistent with the expected value of 0.5 in the vicinity of a critical point.  Analysis of the other PHENIX datasets is currently underway.

\section{Azimuthal Correlations at Low Transverse Momentum}

Critical behavior may also be apparent in the width and shape of correlation functions.  PHENIX has measured azimuthal correlation functions of like-sign pairs at low $p_T$ for several collision systems. The correlations isolate the HBT peak in pseudorapidity by restricting $|\Delta\eta|<0.1$ for each particle pair. Correlations are constructed for low $p_T$ pairs by correlating all particle pairs in an event where both particles lie within the $p_T$ range $0.2<p_{T,1}<0.4$ GeV/c and $0.2<p_{T,2}<0.4$ GeV/c. Note that there is no trigger particle in this analysis. The correlation functions are constructed using mixed events as follows: $C(\Delta\phi) = \frac{dN/d\phi_{data}}{dN/d\phi_{mixed}}\frac{N_{events,mixed}}{N_{events,data}}$. Confirmation of the HBT peak has been made by observing its disappearance in unlike-sign pair correlations and by observing $Q_{invariant}$ peaks when selecting this region.

\begin{figure}[htb]
                 \resizebox{0.75\textwidth}{!}{%
                 \includegraphics{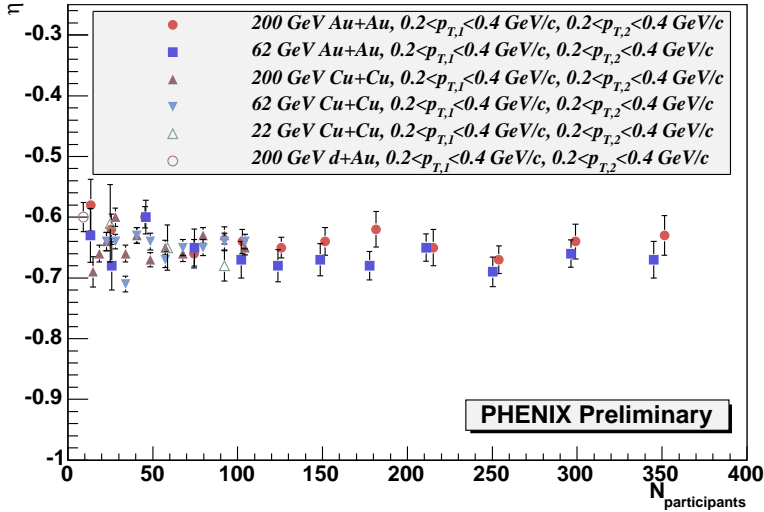}
}
\caption[]{The exponent $\eta$ with d=3 extracted from the like-sign correlation functions as a function of $N_{part}$.}
\label{fig:etaVsNpart}
\end{figure}

Azimuthal correlation functions can be described by a power law function with exponent $\eta$: $C(\Delta \phi) \propto \Delta\phi^{-(d-2+\eta)}$, where d is the dimensionality of the system \cite{Stanley}.  For all collision systems, including 200 GeV d+Au, the extracted value of the exponent $\eta$ is shown in Fig. \ref{fig:etaVsNpart}.  The value of $\eta$ lies between -0.6 and -0.7 with d=3, independent of centrality. Since $\eta$ is constant in heavy ion collisions, does not differ from the d+Au system, and has a value that significantly differs from expectations from a QCD phase transition (e.g. $\eta$=+0.5 for the 3-D Ising model universality class \cite{Reiger}), it is unlikely that critical behavior is being observed in the correlation functions measured thus far.

\begin{figure}[htb]
                 \resizebox{0.75\textwidth}{!}{%
                 \includegraphics{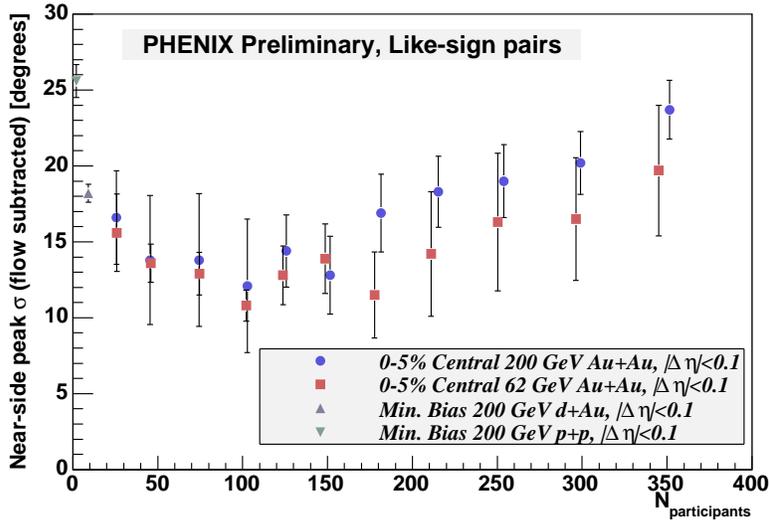}
}
\caption[]{The standard deviation of a Gaussian fit to the HBT peak in like-sign correlation functions as a function of $N_{part}$ for several collision species.}
\label{fig:sigmaVsNpart}
\end{figure}

Near the critical point, it is also expected that the correlation length will diverge with a power law behavior. The HBT peak of the correlation functions with the estimated contribution from elliptic flow subtracted have been fit to a Gaussian distribution. The standard deviation from the fit is shown in Figure \ref{fig:sigmaVsNpart} for several collision species. There is no significant change in the correlation widths between 200 GeV Au+Au and 62.4 GeV Au+Au collisions.

\section{Conclusions}

The fluctuation and correlation measures presented here do not provide a significant indication of the existence of a critical point or phase transition. This does not rule out the possibility that the critical point exists. Further searches will be facilitated by the upcoming RHIC low energy program.


\begin{thebibliography}{99}

 \bibitem{QCDstephanov}
  M.~A.~Stephanov, K.~Rajagopal and E.~V.~Shuryak,
  Phys.\ Rev.\ Lett.\  {\bf 81}, 4816 (1998).
 
\bibitem{Stanley}
  H.~Stanley, {\it Introduction to Phase Transitions and Critical Phenomena} (Oxford, New York and Oxford) 1971.

\bibitem{ua5}
  G.~J.~Alner {\it et al.}  [UA5 Collaboration],
  Phys.\ Rept.\  {\bf 154}, 247 (1987).

\bibitem{e802MF}
  T.~Abbott {\it et al.}  [E-802 Collaboration],
  Phys.\ Rev.\  C {\bf 52}, 2663 (1995).

\bibitem{HIJING}
  X.~N.~Wang and M.~Gyulassy,
  Phys.\ Rev.\  D {\bf 44}, 3501 (1991).

\bibitem{phxMult}
  S.~S.~Adler {\it et al.}  [PHENIX Collaboration],
  Phys.\ Rev.\  C {\bf 71}, 034908 (2005)
  [Erratum-ibid.\  C {\bf 71}, 049901 (2005)].

\bibitem{heiselReview}
  H.~Heiselberg,
  Phys.\ Rept.\  {\bf 351}, 161 (2001).

\bibitem{na49MF}
  C.~Alt {\it et al.}  [NA49 Collaboration],
  Phys.\ Rev.\  C {\bf 75}, 064904 (2007).

\bibitem{clanOriginal}
  A.~Giovannini and L.~Van Hove,
  Z.\ Phys.\  C {\bf 30}, 391 (1986).

\bibitem{emcClan}
  M.~Arneodo {\it et al.}  [European Muon Collaboration],
  Z.\ Phys.\  C {\bf 35}, 335 (1987)
  [Erratum-ibid.\  C {\bf 36}, 512 (1987)].

\bibitem{na22Clan}
  M.~Adamus {\it et al.}  [EHS/NA22 Collaboration],
  Z.\ Phys.\  C {\bf 37}, 215 (1988).

\bibitem{na35Clan}
  J.~Bachler {\it et al.}  [NA35 Collaboration],
  Z.\ Phys.\  C {\bf 57}, 541 (1993).

\bibitem{phxPtFluc}
  S.~S.~Adler {\it et al.}  [PHENIX Collaboration],
  Phys.\ Rev.\ Lett.\  {\bf 93}, 092301 (2004).

\bibitem{phxFlow}
  A.~Adare {\it et al.}  [PHENIX Collaboration],
  Phys.\ Rev.\ Lett.\  {\bf 98}, 162301 (2007).

\bibitem{ceresPTF}
  D.~Adamova {\it et al.}  [CERES Collaboration],
  Nucl.\ Phys.\  A {\bf 727}, 97 (2003).

\bibitem{Korus}
  R.~Korus, S.~Mrowczynski, M.~Rybczynski and Z.~Wlodarczyk,
  Phys.\ Rev.\  C {\bf 64}, 054908 (2001).

\bibitem{Reiger}
  H.~Reiger,
  Phys.\ Rev.\  B {\bf 52}, 6659 (1995)

\bibitem{Csorgo05}
  T.~Csorgo {\it et al.}
  Acta. Phys. Pol. {\bf B36}, 329 (2005).
 

\end{thebibliography}
\end{document}